\documentclass{llncs}
\usepackage[utf8]{inputenc}
\usepackage[IL2]{fontenc}
\usepackage{color}
\usepackage{url}
\usepackage{xspace}
\usepackage{graphicx}
\usepackage{listings}

\newcommand{\Stanse}{\textsc{Stanse}\xspace}
\newcommand{\TC}{\textsc{ThreadChecker}\xspace}
\newcommand{\AC}{\textsc{AutomatonChecker}\xspace}
\newcommand{\RC}{\textsc{ReachabilityChecker}\xspace}
\newcommand{\LC}{\textsc{LockChecker}\xspace}

\begin{document}


\title{{STANSE}: Bug-finding Framework\\ for C Programs}

\author{Jan~Obdr\v{z}\'alek, Ji\v{r}\'\i~Slab\'y and Marek~Trt\'{\i}k}
\institute{Masaryk University, Brno, Czech Republic\\ \email{\{obdrzalek,slaby,trtik\}@fi.muni.cz}}

\maketitle

\begin{abstract}
  \Stanse is a free (available under the GPLv2 license) modular framework
  for finding bugs in C programs using static analysis. Its two main design
  goals are 1) ability to process large software projects like the Linux kernel and
  2) extensibility with new bug-finding techniques with a minimal effort.
  Currently there are four bug-finding algorithms implemented within
  \Stanse: \AC checks properties described in an automata-based formalism,
  \TC detects deadlocks among multiple threads, \LC finds locking errors based
  on statistics, and \RC looks for
  unreachable code. \Stanse has been tested on the Linux kernel, where it
  has found dozens of previously undiscovered bugs.
\end{abstract}



\section{Introduction}
\label{sec:introduction}

During the last decade, bug-finding techniques based on static analysis have
finally come of age. One of the papers to really stir interest
was~\cite{ECCH00}, showing that static analysis can efficiently find many
interesting bugs in real-world code.  This work eventually led to a
successful commercial tool called \textsc{Coverity}~\cite{coverity}. Over the
years, several other successful tools, like
\textsc{CodeSonar}~\cite{codesonar} or \textsc{Klocwork}~\cite{klocwork},
appeared.
However, such fully-featured tools are neither free to obtain, nor is their
code available (e.g.~for developing new algorithms or tailoring the existing
tools to specific tasks). The existing free tools are usually severely
limited in what they can do (e.g.~\textsc{Uno}~\cite{uno},
\textsc{Sparse}~\cite{sparse}, \textsc{Smatch}~\cite{smatch}). One notable
exception is \textsc{FindBugs}~\cite{HP04,findbugs}, a successful tool
working on Java code. \Stanse is intended to fill this gap for the C language. It
can be seen in two ways:
\begin{enumerate}
\item \Stanse is a robust framework (written predominantly in Java) for
  implementing diverse static analysis algorithms. An implemented algorithm
  can be immediately evaluated on large real-world software projects written
  in C as the framework is capable to process such projects (for example, it
  can process the whole Linux kernel). An implementation of such an algorithm
  within the framework is called a \emph{checker}.

  \item As \Stanse already contains four checkers, it can be also seen as a
  working static analysis tool.
\end{enumerate}


The paper is structured as follows. Section~\ref{sec:framework}
describes the functionality provided by the framework, while
Section~\ref{sec:checkers} is devoted to the four existing checkers.
In Section~\ref{sec:results} we present some results of running \Stanse on the Linux
kernel. The last section summarises the basic strengths of \Stanse and
mentions some directions of future development.


\section{Framework Functionality}
\label{sec:framework}

\begin{figure}[!htb]
  \centering
    \includegraphics[width=10cm]{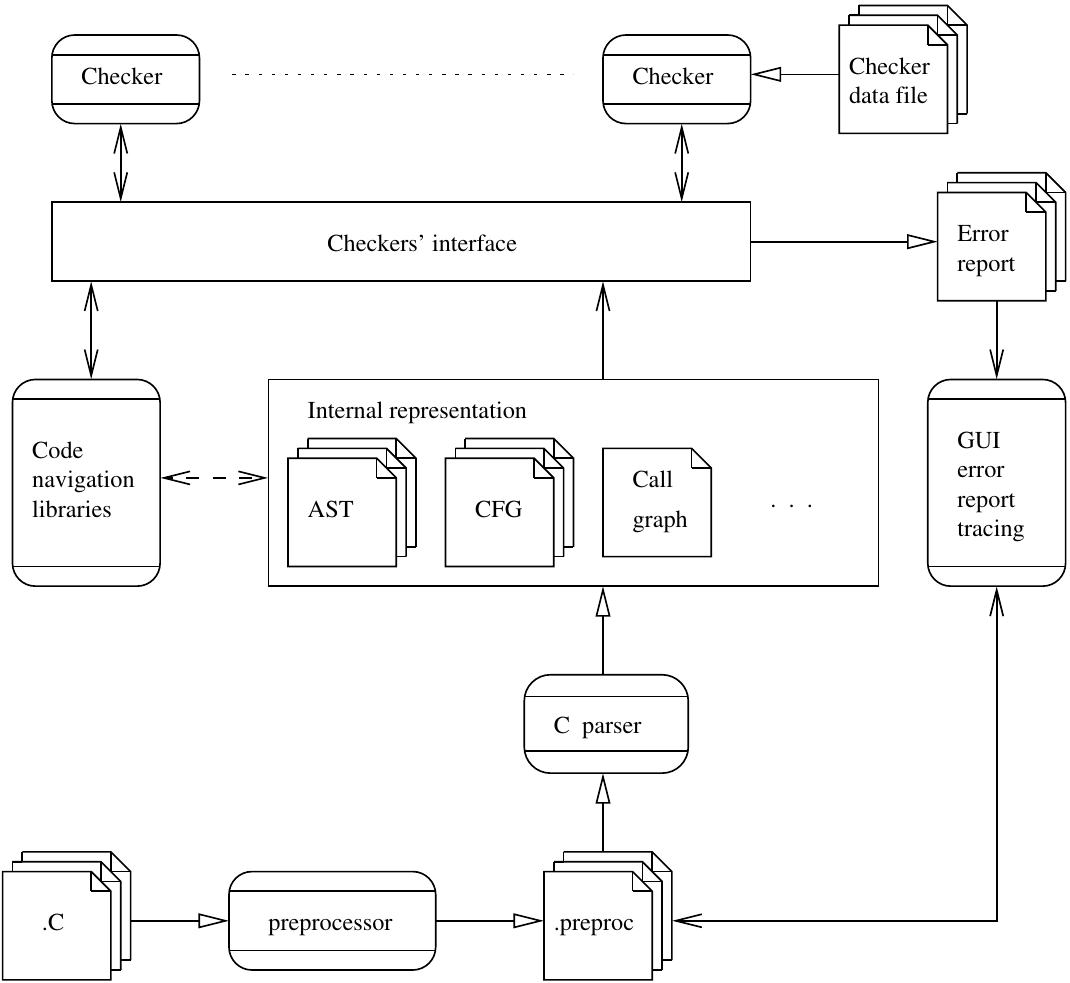}
  \caption{\Stanse framework architecture.}
  \label{fig:architecture}
\end{figure}

The \Stanse framework is modular and fully open. It is designed to allow
static analysis of large software projects like Linux kernel. Furthermore it is
aimed to reduce effort when implementing a new static analysis technique.
Architecture of the framework is depicted in Fig.~\ref{fig:architecture},
and is more or less standard. The bug-finding algorithms are implemented as
\emph{checkers}, and will be described in more detail in
Section~\ref{sec:checkers}. In this section we focus on the functionality of
the framework itself, describing only the non-standard or for some other
reason interesting features.

\subsection{Configuration}

The \Stanse framework contains data structures capturing a configuration of an
analysis to be performed. We always need to know what source files to analyse
and by what checkers. That information we call \emph{configuration}. 

There are several ways how to tell \Stanse which source files to analyse. Besides
the standard possibilities (a given file, all files from a given
directory, all files listed in a list), \Stanse can also derive all
the necessary information from a project \textsf{Makefile}. In this case,
\Stanse also remembers compiler flags for preprocessing purposes. This
functionality is inspired by the \textsc{Sparse}~\cite{sparse} tool. 

The user must also specify the checkers which should be run on the
configured source files. 
There can be many checkers running simultaneously in \Stanse. However they
cannot share any data and proceed independently.
Checkers themselves can also be configured through their own configuration
files. 
The configuration can be passed to the \Stanse framework either via command line
arguments, or using the graphical interface.

\subsection{Parsing Source Files}

The \Stanse framework can process source files written in C, more
specifically in the ISO/ANSI C99 standard together with most of the GNU C
extensions. This allows \Stanse to process software projects like the Linux
kernel. We are currently working on support of other languages, in
particular C++, a prototype implementation for which is included in the
distribution. The parsing pipeline, including a preprocessing the source
files by the standard GNU C preprocessor, is depicted at the bottom in
Fig.~\ref{fig:architecture}.  It is important to note that we do not parse
the source files in a sequence one by one as they appear in a
configuration. Since we use streaming (which we discuss later), the pipeline
is applied as needed for each individual source file.

The parser used in \Stanse is generated using the \textsc{AntLR} tool from
our own annotated C grammar. The reason for us to write our own parser was
that at the time we started to develop \textsc{Stanse} we could not find
free parser which, while being suitable for our purposes, would be able to
parse most of the GNU C. (Linux Kernel makes heavy use of GNU C language
extensions.)  There is one notable exception: \textsc{Clang}, which is
slowly improving and will be able to handle the Linux kernel in the near
future. However, in its current form it still contains bugs and cannot be
used reliably. Our plan is to switch from our parser to \textsc{Clang} once
it becomes stable and feature complete.

Also, one may object that we could use the parser from the GNU compiler
(GCC).  Unfortunately this parser, and most importantly its internal
structures, is not suitable for our purposes.  For example the CFG is built
on the top of RTL or tree representation (we encourage the reader to look
into the GCC manual where these are described).

\subsection{Program Internal Representation}

Once the code is parsed, it is represented using \Stanse's \emph{internal structures}: a call graph
among functions, a control flow graph (CFG) for each function, and an abstract
syntax tree (AST) for the whole file. Subtrees in the AST are referenced from
appropriate CFG nodes. We show these structures in the middle of
Fig.~\ref{fig:architecture}. All these structures can be dumped in a textual
or graphical form.

Since we aim at large software projects consisting of hundreds or thousands of
modules (compilation units), it is often impossible in practice to store the
corresponding internal structures in the memory all at the same time. The \Stanse
framework therefore applies automatic \emph{streaming} of the internal
structures. This is currently performed on a module basis. Instead of parsing
all source modules in the beginning, a module is streamed in only when \Stanse
needs to access some internal structure belonging to the module. In other
words, the internal structures are constructed on demand, in a lazy manner.

If the memory occupied by internal structures exceeds a given limit, some
internal structures have to be freed before another module is streamed
in. The structures to be freed are selected using the LRU (least recently used)
approach: \Stanse discards all internal structures of the module whose
structures are not accessed for the longest time. If the discarded structure
is accessed again later, the corresponding module is streamed back in. Both
laziness of internal structures and streaming are completely invisible to
checkers. 

In the current implementation of streaming, each source file streamed out
from the memory is completely discarded. \Stanse does not back up already
parsed internal structures into auxiliary files before discarding. As a
consequence, when internal structures of the discarded file are needed again,
\Stanse starts the parsing pipeline of the file from scratch to recreate
requested internal structures. Although loading of previously parsed internal
structures from auxiliary files would speed up the streaming process,
profiling of \Stanse's performance on the Linux kernel has not shown
streaming to be a performance bottleneck. However this could be easily
changed if streaming performance becomes a problem in the future.

\subsection{Pointer Analysis in \Stanse}

Since C programs tend to heavily use pointers, it almost always becomes a
necessity to use some form of pointer analysis. There are many different
known approaches to pointer analysis, differing in speed and accuracy. As
each bug-finding/program analysis technique may have different requirements
regarding pointer analysis, a framework like \Stanse
should ideally implement several different pointer analysis techniques and
provide them to its checkers.

Nevertheless, the \Stanse framework currently provides just two pointer
analyses: \textit{Steensgaard}'s~\cite{SG96} and
\textit{Shapiro-Horowitz}'s~\cite{SH97}. Both analyses are \emph{may} analyses
-- they compute an over-approximation of an accurate solution. The
\textit{Steensgard}'s analysis is very fast and it is widely used in practice.
On the other hand it is not very accurate. The \textit{Shapiro-Horowitz}'s
analysis allows parametrisation between \textit{Steensgard}'s and
\textit{Andersen}'s analyses. One can therefore balance between speed of
\textit{Steensgard}'s analysis and accuracy of \textit{Andersen}'s one.

\subsection{Matching Language Constructs}

Many static analyses change their internal state only on some subset of
program expressions. For example, when finding race conditions in a
parallel program, one may only focus on expressions involving synchronisation,
while ignoring all others. The \Stanse framework therefore provides a
specification language for determining a set of program expressions.

The language defines a collection of \emph{patterns}. Each pattern is supposed
to identify a single specific kind of sub-trees in AST of analysed program. A
pattern itself is therefore also a sub-tree of AST, where some of its vertices
are ``special''. They allow to define a set of possible sub-trees at that
vertex.

This is, however, not the only possible approach. For example, in the
METAL~\cite{METAL} specification language, a \texttt{C} expression can be
directly parametrised to define a set. The solution we implemented exploits
the fact that checkers in \Stanse work with AST intensively, and therefore
identifying expressions in terms of AST is very practical.

\subsection{Traversing Internal Representation}
\label{ssec:travers}

Although a checker may need to work with the internal structures in an
arbitrary way, most checkers walk through CFGs using some standard
strategy. To prevent unnecessary reimplementations, the most important and
heavily used traversal methods are implemented directly inside the framework.
With this functionality, one can implement a new checker (or its part) by
specifying
\begin{itemize}
\item whether it should go through CFGs forwards or backwards, breadth-first
  or depth-first,
\item whether the interprocedural walk-through should be performed or not (if
  not, the function calls are ignored), and
\item a method (callback) to be called for each visited node in a CFG.
\end{itemize}
This makes implementation of new algorithms extremely simple.

For example, when a checker needs to implement a forward flow-sensitive analysis,
it may ask the \Stanse framework to traverse paths in CFGs in forward
depth-first manner. This can be implemented by a single call to a
function \texttt{traverseCFGToDepthForward}, which takes as an argument a
CFG and a subclass of \Stanse class \texttt{CFGPathVisitor}. In this class
the checker defines the action which should be taken whenever a CFG
node is visited (already in the requested order). The checker implements the
action in a method \texttt{visit} of the subclass.

In addition, for those interprocedural analyses which do not construct
summaries \Stanse provides an automated traversal among different CFGs
according to function calls (involving automated parameters passing and
value returning). Again, this can be done using a single call to the \Stanse
framework.  

The functionality described in this section is shown in Fig.~\ref{fig:architecture} as
``Code navigating libraries''.


\subsection{Support for Function Summaries}
\label{ssec:summaries}

Interprocedural analyses typically build function summaries. Unfortunately,
these summaries may differ from one analysis to another. Nevertheless, quite
common part in building many summaries is passing formal and actual parameters
to call sites and mapping return values to appropriate variables. Therefore,
the \Stanse framework provides classes simplifying the parameter passing and
values returning for checkers. These classes also provide a conversion of a
given expression in the caller into an equal expression in the called
function. The conversion can also be required in the opposite direction, i.e.
for returned values.

\subsection{The Concept of Checkers}

In the \Stanse framework a checker is an implementation of some concrete
static analysis technique. Each checker has an access to a shared internal
structure of analysed source files. They are also provided with an access to
the algorithms providing navigation in those structures. This is done by an
interface between checkers and internal structure and libraries of the
framework. The interface is depicted in Fig.~\ref{fig:architecture} right
bellow the checkers.

The checkers are integrated in the framework of \Stanse using concrete factory
design pattern. Therefore, to insert a new checker to the framework one needs
to implement generic checker interface and register it to the checkers'
factory of the framework. Then it gains a full access to the features of the
framework accessible through the discussed interface.

It is very easy to integrate a new checker into \Stanse. The process requires
only three simple steps to be fully functional. The first step is to create a
subclass of \Stanse abstract class \texttt{Checker}, say \texttt{MyChecker}.
The most important method to implement is \texttt{check}. There the analysis
algorithm should be implemented.

The second step is to integrate the newly created class into the framework.
This means implementation of \texttt{MyCheckerCreator}, a subclass of
\texttt{CheckerCreator} abstract class. And the final step is to register the
class \texttt{MyCheckerCreator}. It comprises adding a line
\texttt{registerCheckerCreator(new MyCheckerCreator())} at the end of
\texttt{CheckerFactory.java}.

\subsection{Processing Errors}

\begin{figure}[t]
  \centering
    \includegraphics[width=9cm]{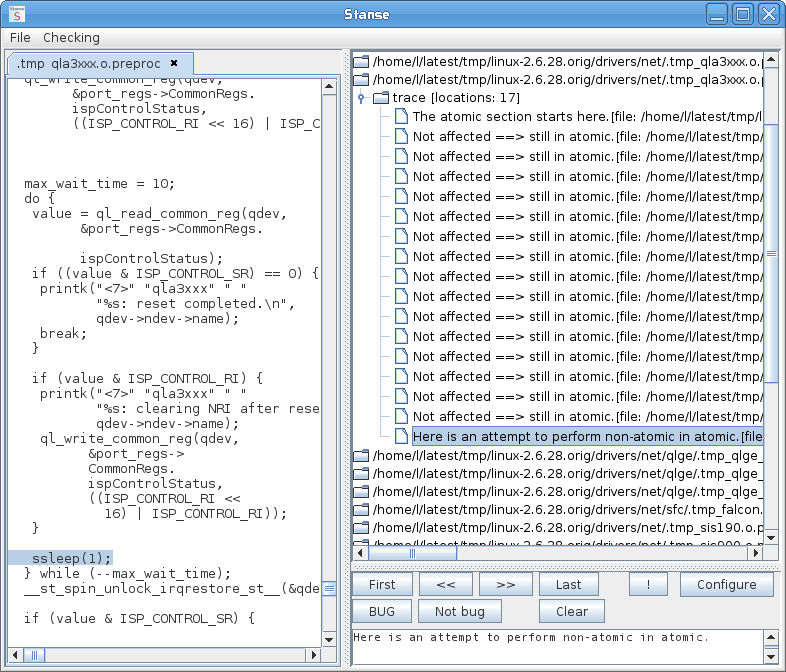}
  \caption{Error trace browser in the \Stanse GUI.}
  \label{fig:bugtrace}
\end{figure}

Once a checker finds an error, it reports the error back to the framework in
the form of an annotated \emph{error trace} - a path in the analysed code demonstrating this
error. A datatype is provided in the \Stanse framework to describe an error. 
In the framework there are then several possibilities how to present the error
traces back to the user of \Stanse: they can be printed to the console,
displayed using a built-in error trace browser in the GUI (see
Fig.~\ref{fig:bugtrace}), or saved to an external file in XML format. This
XML file has a wide variety of possible applications. For example, we supply a
tool transforming the XML file into an \textsc{SQLite} database. The database
is supplemented with a web interface allowing to browse errors in the database
via a web browser. Using the web browser or the built-in graphical error
browser, one can mark errors as real bugs or false positives. \Stanse also
provides various statistics of errors like number of errors per checker,
frequency of errors of the same kind, percentage of false positives (based on
user feedback).

Error reporting and tracing pipeline is depicted to the right of interface and
internal representation of Fig.~\ref{fig:architecture}.


\section{Checkers}\label{sec:checkers}

In this section we briefly describe the four currently available checkers. All
four checkers  are provided with sample configuration so that they can be used
instantly, however they can be configured differently when necessary. 

\subsubsection{\AC}
is heavily influenced by~\cite{ECCH00}. It takes, as an input, a set of
finite-state automata that describe the properties we want to check,
patterns which match against the code to be checked, and finally
transitions, i.e. pairing of patterns and
automaton state changes. Properties like locking discipline, interrupt
management, null pointer dereference, dangling pointers and many others can be
described this way.

\begin{figure}[!htb]
  \centering
    \includegraphics[width=3cm]{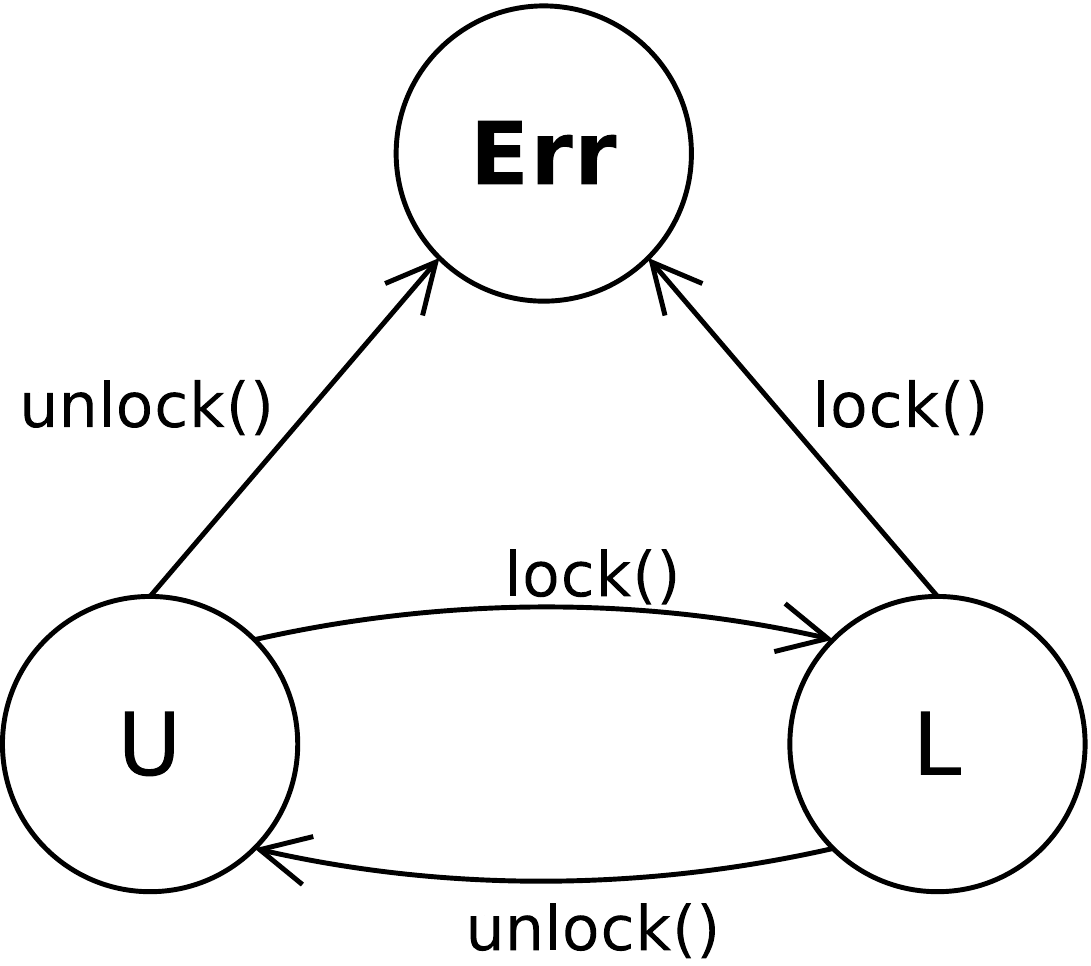}
  \caption{Automaton for locks checking}
  \label{fig:automaton}
\end{figure}

An example of the locking checker is presented in Fig.~\ref{fig:automaton}.
The automaton starts in the unlocked state (\textsf{U}) and a transition is
made when there is an outgoing edge from the current state with a pattern
matching the action currently performed by the analysed program. E.g. if
there is an unlock action while the automaton is in an unlocked state, an error
is reported.

Compared to the implementation described in~\cite{ECCH00} and~\cite{HCXE02},
our technique differs in several aspects. In particular, we do not use
metacompilation, automata are not input-language specific (a pattern
matching is used instead), and the interprocedural analysis is done in the
context of a single input file.

\subsubsection{\LC}
accounts statistics about variable accesses. It also tracks which locks are
locked while each of the variable is accessed. Again both variable accesses
and locks are specified by patterns.

Then, combining the information about accesses and locks held, it
counts a statistics in how many cases each variable is accessed while some
lock is held. If the difference is proportional, an error is reported. So if,
for instance, some variable is changed 99 times while some lock is held and in
one case the lock is not, this is reported as a possible error. The boundary
is currently set to 70, so that at least 70\% of accesses must be under locks.
The rest (30\%) is then reported. This work is based upon~\cite{lockchecker}.

\subsubsection{\TC}
aims to check for possible deadlocks in concurrent programs. The technique
is based on the notions of locksets of~\cite{relay} and deadlock detection
by looking for cycles in \emph{resource allocation graphs (RAGs)}. \TC first
tries to identify the parts of the code which can run in parallel, as
different threads. This is performed by searching of functions instantiating
threads (such as \texttt{pthread\_create}). Or, for the Linux kernel, we also
specify manually which hooks may be run parallel.

Then the checker builds a set of \emph{dependency graphs} for each such
thread. A dependency graph statically represents possible locksets during
one execution of a thread. Dependency graphs are then combined and transformed
into RAGs. If there is a circular lock dependency, RAG contains a cycle. In
such case an error is reported to the user.

\subsubsection{\RC}
searches CFGs for unreachable nodes. These are then reported as warnings or
errors, depending on importance (e.g. superfluous semicolons are less
important than unused branch). The primary goal of \RC is to demonstrate the
simplicity of a new checker implementation. With a help of the framework
features described in Subsection~\ref{ssec:travers}, the code of the checker
has less than 200 lines including the mentioned error/warning classification
and many strings and comments.

Even though it is a very simple checker it was still able to find serious bugs
in the kernel. For example a superfluous semicolon can cause unexpected
unconditional returns from functions like in the following code: \textsf{if
(cond)\underline{;} return;}.


\section{Results on the Linux Kernel}
\label{sec:results}

We have several reasons to choose the Linux kernel for testing \Stanse: the
kernel is a large and freely available codebase, it fully exercises most
of the features of ISO/ANSI C99 and GNU C extensions, it is under constant
development (there is a constant income of new bugs), and the absence of bugs is
of a great concern. 

We applied \Stanse, together with the four checkers described in the
previous section, to the Linux kernel version 2.6.28. The \AC was configured
with three automata describing the following types of errors:
\begin{itemize}
\item incorrect pairing of functions (imbalanced locking, reference
  counting errors)
\item bugs in pointer manipulation (null dereference, dangling pointers, etc.)
\item deadlocks caused by sleeping inside spinlocks or interrupt handlers
\end{itemize}

The running time of \Stanse on a common desktop machine with two 2.5\,GHz
cores and 4\,GiB of memory was under two hours. The memory usage of the Java
process oscillated between 400 and 1000\,MiB. The number of errors found by
the checkers is presented in Table~\ref{tab:results}. Let us note that \RC
actually found 751 errors, but 720 of them are of low importance
(including 696 superfluous semicolons) and they are omitted from our
statistics.

\begin{table*}[tb]
\begin{center}
\setlength{\tabcolsep}{3pt}
\begin{tabular}{|l|l|c|c|c|c|} \hline
\textbf{Checker} & \textbf{Automaton} &
\multicolumn{3}{|c|}{\textbf{Errors}} & \textbf{Real/classified} \\ \cline{3-5}
&&\textbf{Found} & \textbf{Real} & \textbf{False pos.} & \textbf{error ratio}\\ \hline\hline
& Pairing                    & 266 & 65  & 143 & 31.3\,\% \\ \cline{2-6}
\AC & Pointers               & 86  & 48  & 37  & 56.5\,\% \\ \cline{2-6}
& Deadlocks                  & 35  & 16  & 18  & 47.1\,\% \\ \hline
\multicolumn{2}{|l|}{\LC}    & 13  & 6   & 7   & 46.2\,\%  \\ \hline
\multicolumn{2}{|l|}{\TC}    & 20  & 9   & 11  & 45.0\,\% \\ \hline
\multicolumn{2}{|l|}{\RC}    & 31  & 31  & 0   & 100.0\,\%  \\ \hline\hline
\multicolumn{2}{|l|}{\textbf{Overall}}  & \textbf{451} & \textbf{175} &
\textbf{216} & \textbf{47.9\,\%} \\ \hline
\end{tabular}
\end{center}
\caption{\Stanse results on the Linux kernel version 2.6.28.}
\label{tab:results}
\end{table*}

We have manually analysed all the found errors and classified them as real
errors or false positives (with an exception of 60 errors found by \AC where
we are not able to decide in a short time whether it is a false positive or
not). Note that the checkers do not produce any false
negatives (assuming there is no bug in the checkers' implementation). The
reason is that all the checkers
implement may analyses, overapproximating the set of error behaviours.  The numbers of real errors, false
positives and the ratio of real errors to all classified errors can be also
found in Table~\ref{tab:results}. The overall ratio of real errors to all
classified errors is not high: 47.9\%.  However, \Stanse in the current
version does not have any thorough false positive filtering technique, which
may be implemented in future.

More than 70 of the 169 real errors have been reported to kernel developers
and fixed in the following kernel releases (the rest have been independently
discovered and reported by someone else or the incorrect code disappeared
from the kernel before we finished our evaluation of found errors). Some of
the reported bugs remained undiscovered for more than seven years (for
illustration, see our report at \url{http://lkml.org/lkml/2009/3/11/380}).

We have reported another 60 bugs found by \Stanse in the subsequent versions
of the kernel. This number is increasing every month.

\subsection{Important Bugs found by \Stanse}
Although checkers currently implemented in \Stanse are based on widely known
techniques, running them on the Linux kernel helped to uncover several
important bugs. In the text below we present two typical bugs discovered by
\Stanse, each using a different checker.

\subsubsection{AutomatonChecker}
Many bugs found by the \AC trigger only under specific conditions, however
some of them may be visible to the user. Consider this code excerpt taken
from the 2.6.27 kernel, \textsf{drivers/pci/hotplug/pciehp\_core.c} file,
\textsf{set\_lock\_status} function:

\begin{lstlisting}[language=C]
mutex_lock(&slot->ctrl->crit_sect);
/* has it been >1 sec since our last toggle? */
if ((get_seconds() - slot->last_emi_toggle) < 1)
	return -EINVAL;
\end{lstlisting}

Note that the call to \textsf{mutex\_lock} function is followed by an \textbf{if}
statement, which returns immediately in the true branch, omitting a call to 
\textsf{mutex\_unlock}. In fact this
deadlock could be easily triggered by a user. It is sufficient to write "1" to
\textsf{/sys/bus/pci/slots/.../lock} file twice within a second.

\subsubsection{ThreadChecker}
An example of non-trivial error which could not be found by the \AC. The code
described here is from the 2.6.28 kernel, file \textsf{fs/ecryptfs/messaging.c}.

There are three locks in the code,
\textsf{msg\_ctx-\textgreater mux}, which is local per context, and two
global locks --
\textsf{ecryptfs\_daemon\_hash\_mux} and
\textsf{ecryptfs\_msg\_ctx\_lists\_mux}.

Let us denote lock dependencies as a binary relation where the first component
depends on the second. I.e. \textsf{lock(A)} followed by \textsf{lock(B)}
means dependency B on A, and we write $A\leftarrow B$.

\begin{lstlisting}[language=C,numbers=left,escapeinside={XY}{YX}]
int ecryptfs_process_response(...)
{
	...
	mutex_lock(&msg_ctx->mux); XY\label{ecrypt_mux1}YX
	mutex_lock(&ecryptfs_daemon_hash_mux); XY\label{ecrypt_hmux1}YX
	...
	mutex_unlock(&ecryptfs_daemon_hash_mux);
	...
unlock:
	mutex_unlock(&msg_ctx->mux);
out:
	return rc;
}
\end{lstlisting}
Here the two locks on lines~\ref{ecrypt_mux1} and~\ref{ecrypt_hmux1} give
\textsf{msg\_ctx-\textgreater mux} $\leftarrow$
\textsf{ecryptfs\_daemon\_hash\_mux}.
\begin{lstlisting}[language=C,numbers=left,escapeinside={XY}{YX},firstnumber=last]
static int ecryptfs_send_message_locked(...)
{
	...
	mutex_lock(&ecryptfs_msg_ctx_lists_mux); XY\label{ecrypt_lmux2}YX
	...
	mutex_unlock(&ecryptfs_msg_ctx_lists_mux);
	...
}

int ecryptfs_send_message(...)
{
	int rc;

	mutex_lock(&ecryptfs_daemon_hash_mux); XY\label{ecrypt_hmux2}YX
	rc = ecryptfs_send_message_locked(...) XY\label{ecrypt_call}YX
	mutex_unlock(&ecryptfs_daemon_hash_mux);
	return rc;
}
\end{lstlisting}
At line~\ref{ecrypt_call}, function \textsf{ecryptfs\_send\_message\_locked}
is called from \textsf{ecryptfs\_send\_message}, hence the locks at
lines~\ref{ecrypt_lmux2} and~\ref{ecrypt_hmux2} generate lock dependency of
\textsf{ecryptfs\_daemon\_hash\_mux} $\leftarrow$
\textsf{ecryptfs\_msg\_ctx\_lists\_mux}.
\begin{lstlisting}[language=C,numbers=left,escapeinside={XY}{YX},firstnumber=last]
int ecryptfs_wait_for_response(...)
{
	...
	mutex_lock(&ecryptfs_msg_ctx_lists_mux);
	mutex_lock(&msg_ctx->mux);
	...
	mutex_unlock(&msg_ctx->mux);
	mutex_unlock(&ecryptfs_msg_ctx_lists_mux);
	return rc;
}
\end{lstlisting}
Finally, this function introduces \textsf{ecryptfs\_msg\_ctx\_lists\_mux}
$\leftarrow$ \textsf{msg\_ctx-\textgreater mux}.

Composing these results together the following circular dependency of these
three locks was found:
\begin{itemize}
\item \textsf{msg\_ctx-\textgreater mux} $\leftarrow$ \textsf{ecryptfs\_daemon\_hash\_mux}
\item \textsf{ecryptfs\_daemon\_hash\_mux} $\leftarrow$ \textsf{ecryptfs\_msg\_ctx\_lists\_mux}
\item \textsf{ecryptfs\_msg\_ctx\_lists\_mux} $\leftarrow$
  \textsf{msg\_ctx-\textgreater mux}
\end{itemize}

This issue was later confirmed as a real bug leading to a
deadlock\footnote{http://lkml.org/lkml/2009/4/14/527}.


\section{Conclusions and Future Work}

\Stanse is a free Java-based framework design for simple and efficient implementation of
bug-finding algorithms based on static analysis. The framework can process
large-scale software projects written in ISO/ANSI C99, together with most
the GNU C extensions. \Stanse does not
currently use any new techniques -- its novelty comes from the fact that (to
our best knowledge) there is no other open-source framework with comparable
applicability and efficiency. We note that more than 130 bugs found by
\Stanse have been reported to and confirmed by Linux kernel developers
already. More information and the tool itself can be found at
\url{http://stanse.fi.muni.cz/}.

\subsubsection{Future Work}
We plan to improve the framework in several directions. Firstly we are currently
working on C++ support. Furthermore we plan to provide \Stanse in the form of
an IDE plug-in, e.g. for Eclipse and NetBeans. A lot of work can be done in the area
of automatic false alarm filtering and error importance classification.
Independently of developing new features, we would like to speed up the
framework as well. To this end we intend to replace the current parser
written in Java by an optimised parser written in C, to replace the XML
format of internal structures by a more succinct representation, to add a
support for function summaries, etc.

\subsubsection{Acknowledgements}
\label{sec:acknowledgements}

Jan Ku\v{c}era is the author of the \TC. We would like to thank Linux kernel
developers, and Cyrill Gorcunov for \Stanse alpha testing and useful
suggestions. All authors are supported by the research centre Institute for
Theoretical Computer Science (ITI), project No. 1M0545.


\bibliographystyle{plain}
\bibliography{stanse}

\end{document}